# Thermally induced all-optical ferromagnetic resonance in thin YIG films

E. Schmoranzerová[1]*, J. Kimák[1], R. Schlitz[3], S.T. B. Goennenwein[3,6], D. Kriegner[2,3], H. Reichlová[2,3], Z. Šobáň[2], G. Jakob[5], E.-J. Guo[5], M. Kläui[5], M. Münzenberg[4], P. Němec[1], T. Ostatnický[1]

[1]Faculty of Mathematics and Physics, Charles University, Prague, 12116, Czech Republic

[2]Institute of Physics ASCR v.v.i , Prague, 162 53, Czech Republic

[3]Technical University Dresden, 01062 Dresden, Germany

[4]Institute of Physics, Ernst-Moritz-Arndt University, 17489, Greifswald, Germany

[5]Institute of Physics, Johannes Gutenberg University Mainz, 55099 Mainz, Germany

[6] Department of Physics, University of Konstanz, 78457 Konstanz, Germany

Laser-induced magnetization dynamics is one of the key methods of modern opto-spintronics which aims at increasing the spintronic device speed[1,2]. Various mechanisms of interaction of ultrashort laser pulses with magnetization have been studied, including ultrafast spin-transfer[3], ultrafast demagnetization[4], optical spin transfer and spin orbit torques [5,6,7], or laser-induced phase transitions[8,9]. All these effects can set the magnetic system out of equilibrium, which can result in precession of magnetization. Laser-induced magnetization precession is an important research field of its own as it enables investigating various excitation mechanisms and their ultimate timescales[2]. Importantly, it also represents an all-optical analogy of a ferromagnetic resonance (FMR) experiment, providing valuable information about the fundamental parameters of magnetic materials such as their spin stiffness, magnetic anisotropy or Gilbert damping[10]. The "all-optical FMR" (AO-FMR) is a local and non-invasive method, with spatial resolution given by the laser spot size, which can be focused to the size of few micrometers. This makes it particularly favourable for investigating model spintronic devices.

Magnetization precession has been induced in various classes of materials including ferromagnetic metals[11], semiconductors[10, 12], or even in materials with a more complex spin structure, such as non-collinear antiferromagnets[13]. Ferrimagnetic insulators, with Yttrium Iron Garnet (YIG, $Y_3Fe_5O_{12}$) as the prime representative[14], are of particular importance for spintronic applications owing to their high spin pumping efficiency[15] and the lowest known Gilbert damping[16]. However, inducing magnetization dynamics in ferrimagnetic garnets using optical methods is quite challenging, as it requires large photon energies (bandgap of YIG is $E_g \approx 2.8$ eV)[17]. This spectral region is rather difficult to access with most common ultrafast laser systems, which are usually suited for near-infrared wavelengths. Therefore, methods based mostly on non-thermal effects, such as inverse Faraday[18,19] and Cotton-Mouton effect[20] or photoinduced magnetic anisotropy[21, 22] have been used to trigger the magnetization precession in YIG so far. For these phenomena to occur, large laser fluences of tens of mJ/cm$^2$ are required[23]. In contrast, laser fluences for a thermal excitation of magnetization precession usually do not exceed tens of μJ/cm$^2$ (Refs. 12, 21, 13). Using the low fluence excitation regime allows for the determination of quasi-equilibrium material parameters, not influenced by strong laser pulses. In magnetic garnets, an artificial engineering of the magnetic anisotropy via the inclusion of bismuth was necessary to achieve thermally-induced magnetization precession[21].

In this paper, we show that magnetization precession can be induced thermally by femtosecond laser pulses in a thin film of pure YIG only by adding a metallic capping layer. The laser pulses locally heat the system, which sets the magnetization out of equilibrium due to the temperature dependence of its magnetocrystalline anisotropy. This way we generate a Kittel (n = 0, homogeneous precession) FMR mode, with a precession frequency corresponding to the quasi-equilibrium magnetic anisotropy of the thin YIG film[10]. We thus prove that the AO-FMR method is applicable for determining micromagnetic parameters of thin YIG films. Using the AO-FMR technique we revealed that at low temperature the Kittel mode damping is significantly faster than at room-temperature, in accord with previous FMR experiments[24,25].

Our experiments were performed on a 50 nm thick layer of pure YIG grown by pulsed-laser deposition on a gadolinium-gallium-garnet (GGG) (111)-oriented substrate. One part of the film was covered by 8 nm of Au capping layer, the other part by Pt capping, both being prepared by ion-beam sputtering. Part of the sample was left uncapped as a reference. X-ray diffraction confirmed the excellent crystal quality of the YIG film with a very low level of growth-induced strain, as described in detail in Ref. 26. The magnetic properties were further characterized using SQUID magnetometry and ferromagnetic resonance experiments, showing the in-plane orientation of magnetization (see Supplementary Material, Part 1 and Figs. S1 and S2). The deduced low-temperature (20 K) saturation magnetization $\mu_0 M_s \approx 180$ mT is in agreement with results published on qualitatively similar samples[27] again confirming a good quality of the studied YIG film. Magnetic anisotropy of the system at 20 K was established from an independent magneto-optical experiment (Ref. 28), the corresponding anisotropy constants for cubic anisotropy of the first and second order are $K_{c1} = -4680$ J/m$^3$ and $K_{c2} = -223$ J/m$^3$, while the overall uniaxial out-of-plane anisotropy is vanishingly small.

Laser-induced dynamics was studied in a time-resolved magneto-optical experiment in transmission geometry, as schematically shown in Fig. 1(a). An output of a Ti:Sapphire oscillator generating 200 fs laser pulses was divided into a strong pump beam, with fluences tuned between 70 and 280 μJ/cm$^2$, and a 20-times weaker probe beam. The beams were focused on a 30 μm spot on the sample, which was placed in a cryostat and kept at cryogenic temperatures (typically 20 K). An external magnetic field (up to 550 mT) generated by an electromagnet was applied in *y* direction (see Fig. 1). The wavelength of pump pulses (800 nm) was set well below the absorption edge of the YIG layer, as indicated in the transmission spectrum of the sample in Fig. 1(b). The wavelength of probe pulses (400 nm) was tuned to match the maximum of the magneto-optical response of bulk YIG [see inset in Fig. 1(b) and Ref. 29].

The detected time-resolved magnetooptical (TRMO) signal corresponding to the rotation of polarization plane of the probe beam Δβ, was measured as a function of the time delay Δt between pump and probe pulses. In Fig. 1(c), we show an example of TRMO signals observed in uncapped YIG and two YIG/metal heterostructures. Clearly, in the presence of the metallic capping layer an oscillatory TRMO signal is observed, whose amplitude depends on the capping metal used. Frequency and damping of the oscillations, on the other hand, remain virtually unaffected by the type of the capping layer, while no oscillations are observed in the uncapped YIG sample.

The TRMO signals can be phenomenologically described by a damped harmonic function after removing a slowly varying background (see Supplementary Material, Part 2 and Fig. S3),[12]

$$\Delta\beta(\Delta t) = A\cos(2\pi f \Delta t + \varphi)\exp(-\Delta t/\tau), \quad (1)$$

where $A$ is the amplitude of precession, $f$ its frequency, $\varphi$ the phase and $\tau$ the damping time. The fits are shown in Fig. 1(c) as solid lines.

In order to demonstrate that the TRMO signals result from (laser-induced) magnetization dynamics, we varied the external magnetic field $H_{ext}$ and extracted the particular precession parameters by fitting the detected signals by Eq. (1). As depicted in Fig. 2(a), the experimentally observed dependence of the precession frequency on the applied field is in excellent agreement with the solution of Landau-Lifshitz-Gilbert (LLG) equation, using the free energy of a [111] oriented cubic crystal [see Supplementary, section 5, Eq. (S5) and Ref. 28]. This correspondence with the LLG model proves that our oscillatory signals reflect indeed the precession of magnetization in uniform (Kittel) mode in YIG. We stress that the precession frequency is inherent to the YIG layer and does not depend on the type of the capping layer.

The detection of the uniform Kittel mode can be further confirmed by comparing the frequency of the oscillatory TRMO signal with the frequency of resonance modes observed in a conventional, microwave-driven ferromagnetic resonance (MW-FMR) experiment. The MW-FMR experiment was performed in the in-plane ($\theta_H = 0°$) and out-of-plane ($\theta_H = 90°$) geometry of the external field. We measured the TRMO signals in YIG/Au sample in a range of magnetic field angles $\theta_H$ and modelled the angular dependency of $f$ by LLG equation with the same parameters that were used in Fig. 2(a). The output of the model is presented in Fig. 2(b), together with precession frequencies obtained from TRMO and FMR experiments. The MW-FMR data fit well to the overall trend, confirming the presence of uniform magnetization precession [Ref. 30]

To find the exact physical mechanism that triggers laser-induced magnetization precession in our YIG/metal bilayers, we measured the TRMO signals at different sample temperatures $T$. For comparison we calculated also the dependence of $f$ on the first order cubic anisotropy constant $K_{c1}$ from the LLG equation, which is shown in the inset of Fig. 2(c). This graph reveals that $f$ should be directly proportional to $K_{c1}$ in the studied range of temperatures. In Fig 2(c) we plot $f$ as a function of $T$, together with the temperature dependence of $K_{c1}(T)$ obtained from Ref. 28 and Ref. 32. Clearly, both $K_{c1}$ and $f$ show a similar trend in temperature. Considering also the temperature dependence of the precession amplitude [see Fig. S5 (a) and Section 4 of Supplementary Material], we identify the pump pulse-induced heating and consequent modification of the magnetocrystalline anisotropy constant $K_{c1}$ as the dominant mechanism driving laser-induced magnetization precession.

In order to estimate the pump-induced increase in quasi-equilibrium temperature of the sample, we first fit the temperature dependence of the parameter $K_{c1}$ reported in literature by a second order polynomial [Fig. 2(c)]. Owing to the linear relation between $f$ and $K_{c1}$ and the known temperature dependence of $f$, the measured dependence of $f$ on pump fluence $I$ can be converted to the intensity dependence of the temperature increase $\Delta T(I)$, which is shown Fig. 2(d). As expected, higher fluence leads to more pronounced heating, which results in a decrease of the precession frequency. Note that for the highest intensity of 300 $\mu J/cm^2$, the sample temperature can increase by almost 80 K.

Nature of the observed laser-induced magnetization precession was further investigated by comparing samples with different capping layers. In Fig. 3(a) we show the amplitude $A$ of the oscillatory signal in the YIG/Pt and YIG/Au layers as a function of $I$. The difference between the samples is apparent both in the absolute amplitude of the precession and in its increase with $I$, the YIG/Pt showing stronger precession. Furthermore, precession damping is stronger in YIG/Au than in YIG/Pt, as apparent from Fig 3 (b) where effective Gilbert damping parameter $\alpha_{eff}$ is presented as a function of $H_{ext}$. These values of $\alpha_{eff}$ were obtained by fitting the TRMO data by the LLG equation, as described in the Supplementary Material (Section 5). Despite the relatively large fitting error, we can still see that YIG/Pt shows slightly lower $\alpha \approx$ 0.020, while the YIG/Au has $\alpha \approx 0.025$. To understand these differences, we modeled the propagation of laser-induced heat in GGG/YIG/Pt and GGG/YIG/Au multilayers by using the heat equation (see Supplementary Material, Section 7). In Fig. 3(c), $\Delta T$ is presented as a function of time delay $\Delta t$ after pump excitation for selected depths from the sample surface. In Fig. 3(d), the same calculations are presented for variable depths and fixed $\Delta t$. The model clearly demonstrates that a significantly higher $\Delta T$ can be expected in the Pt-capped layer simply due to its smaller reflection coefficient as compared to Au-capping (see Supplementary Material, Section 7). This in turn leads to a higher amplitude of the laser-induced magnetization precession in YIG/Pt compared to the YIG/Au, as apparent in Fig. 3(a).

According to our model, an extreme increase in temperature is induced in the first few picoseconds after excitation, which acts as a trigger of magnetization precession. After approximately 10 ps, precession takes place in quasi-equilibrium conditions. The system returns to equilibrium on a timescale of nanoseconds, which shows also in the TRMO signals as the slowly varying background (Fig. S3). The precession frequency we detect reflects the quasi-equilibrium state of the system. Therefore, the temperature increase $\Delta T$ deduced from the TRMO signal can be compared with our model for large time delays after the excitation ($\Delta t >> 10$ *ps)*. In YIG/Au sample, the experimental values of $\Delta T = (25 \pm 10)$ *K* for excitation intensity of *150* $\mu J/cm^2$ [see Fig. 2(d)], while the model gives us $\Delta T \approx 5$K [Fig. 3 (c)]. Clearly, the values match in the order of magnitude but there is a factor of $\approx 5$ difference. This difference results from the boundary conditions of the model that assumes ideal heat transfer between the sample and the holder, which is experimentally realized using a silver glue with less than perfect performance at cryogenic conditions.

From Fig. 3(d) it also follows that large thermal gradients are generated across the 50 nm layer. This could lead to significant inhomogeneity in magnetic properties of the layer, that would increase the damping parameter $\alpha$ by an extrinsic term. In our TRMO measurements, $\alpha$ is indeed very large for a typical YIG sample ($\alpha_{TRMO} \approx 2\text{-}2.5 \times 10^{-2}$) and exceeds the value obtained from room-temperature MW-FMR by almost an order of magnitude ($\alpha_{FMR} \approx 1 \times 10^{-3}$, see Supplementary Material, Section 1b). As the modeled thermal gradient alone cannot account for such a large change in Gilbert damping (see Supplementary Material, Section 6), we attribute this increase in Gilbert damping to the difference in the ambient temperatures. Large change of Gilbert damping (by a factor of 30) between room and cryogenic (20 K) temperature has recently been reported on a seemingly high quality YIG thin film[24]. It was explained in terms of the presence of rare earth or $Fe^{2+}$ impurities that are activated at cryogenic temperatures. It is likely that the same process occurs in our sample. Even though other mechanisms related to the optical excitation can also contribute to the increase in $\alpha_{TRMO}$ (see Supplementary Material, Section 6), the all-optical and standard FMR generated Kittel modes correspond very well [see Fig. 2(b)]. Furthermore, also the observed sample-dependent Gilbert damping is consistent with this explanation. The YIG/Pt sample is heated to higher temperature by the pump laser pulse [Fig. 3(c), (d)] than the YIG/Au sample, which according to Ref. 24 corresponds to a lower Gilbert damping. It is worth noting that damping parameter can be increased also by spin-pumping from YIG to the metallic layer. However, this effect is expected to be significantly higher when Pt is used as a capping, which does not agree with our observations.

In conclusion, we demonstrated the feasibility of the all-optical ferromagnetic resonance method in 50-nm thin films of plain YIG. Magnetization precession can be triggered by laser-induced heating of a metallic capping layer deposited on top of the YIG film. The consequent change of sample temperature modifies its magnetocrystalline anisotropy, which sets the system out of equilibrium and initiates the magnetization precession. Based on the field dependence of precession frequency, we identify the induced magnetization dynamics as the fundamental (Kittel) FMR mode, which is virtually independent of the type of capping and reflects the quasi-equilibrium magnetic anisotropy. The Gilbert damping parameter is influenced by line-broadening mechanism due to low-temperature activation of impurities, which is an important aspect to be taken into account for low-temperature spintronic device applications.

Regarding the efficiency of the optical magnetization precession trigger, it was found that the type of capping layer strongly influences the precession amplitude. The precession in YIG/Pt attained almost twice the amplitude of that in YIG/Au under the same conditions. This indicates that a suitable choice of capping layer should be considered in an optimization of this local non-invasive magnetometric method.

**Acknowledgments:**

This work was supported in part by the INTER-COST grant no. LTC20026 and by the EU FET Open RIA grant no. 766566. We also acknowledge CzechNanoLab project LM2018110 funded by MEYS CR for the financial support of the measurements at LNSM Research Infrastructure and the German Research Foundation (DFG SFB TRR173 Spin+X projects A01 and B02 #268565370).

**FIGURES**

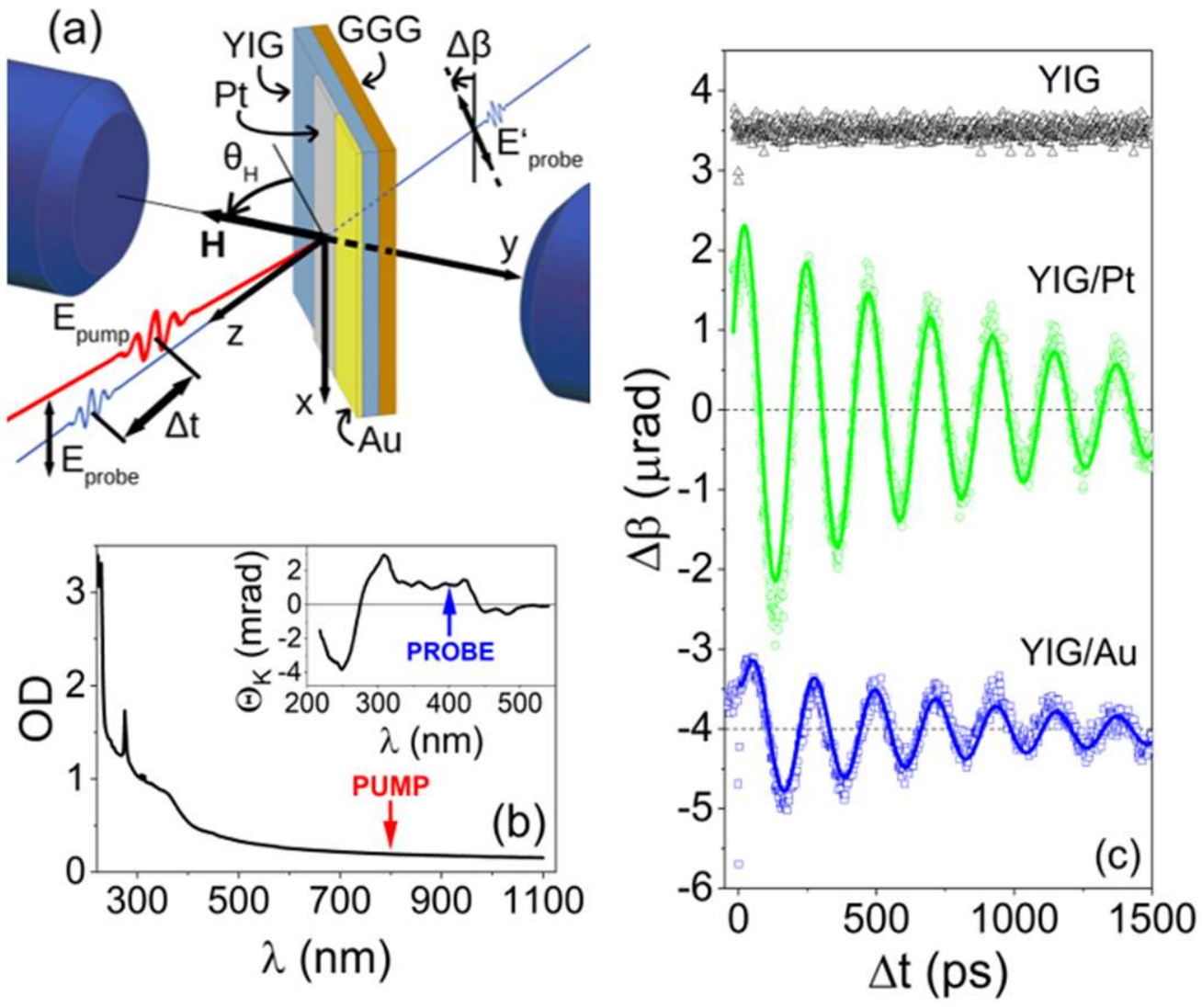

Fig. 1: (a) Schematic illustration of the pump&probe experimental setup, where $E_{probe}$ is the probe beam linear polarization orientation which is rotated by an angle $\Delta\beta$ after transmission through the sample with respect to the orientation $E'_{probe}$. An external magnetic field $H_{ext}$ is applied at an angle $\theta_H$. (b) Absorption spectrum of the studied YIG sample, where OD stands for the optical density defined as minus the decadic logarithm of sample transmittance. The red arrow indicates the wavelength of the pump beam $\lambda_{PUMP}$ = 800 nm. Inset: Spectrum of Kerr rotation $\Theta_K$ of bulk YIG crystal[29]. The blue arrow shows the wavelength of the probe beam $\lambda_{PROBE}$ = 400 nm. (c) Typical time-resolved magneto-optical signals of a plain 50 nm YIG film (black dots), YIG /Pt (green dots) and YIG/Au bilayer (blue dots) at 20 K and $\mu_0 H_{ext}$ = 100 mT, applied at an angle $\theta_H$ = 40°. Lines indicate fits by Eq. (1). The data were offset for clarity.

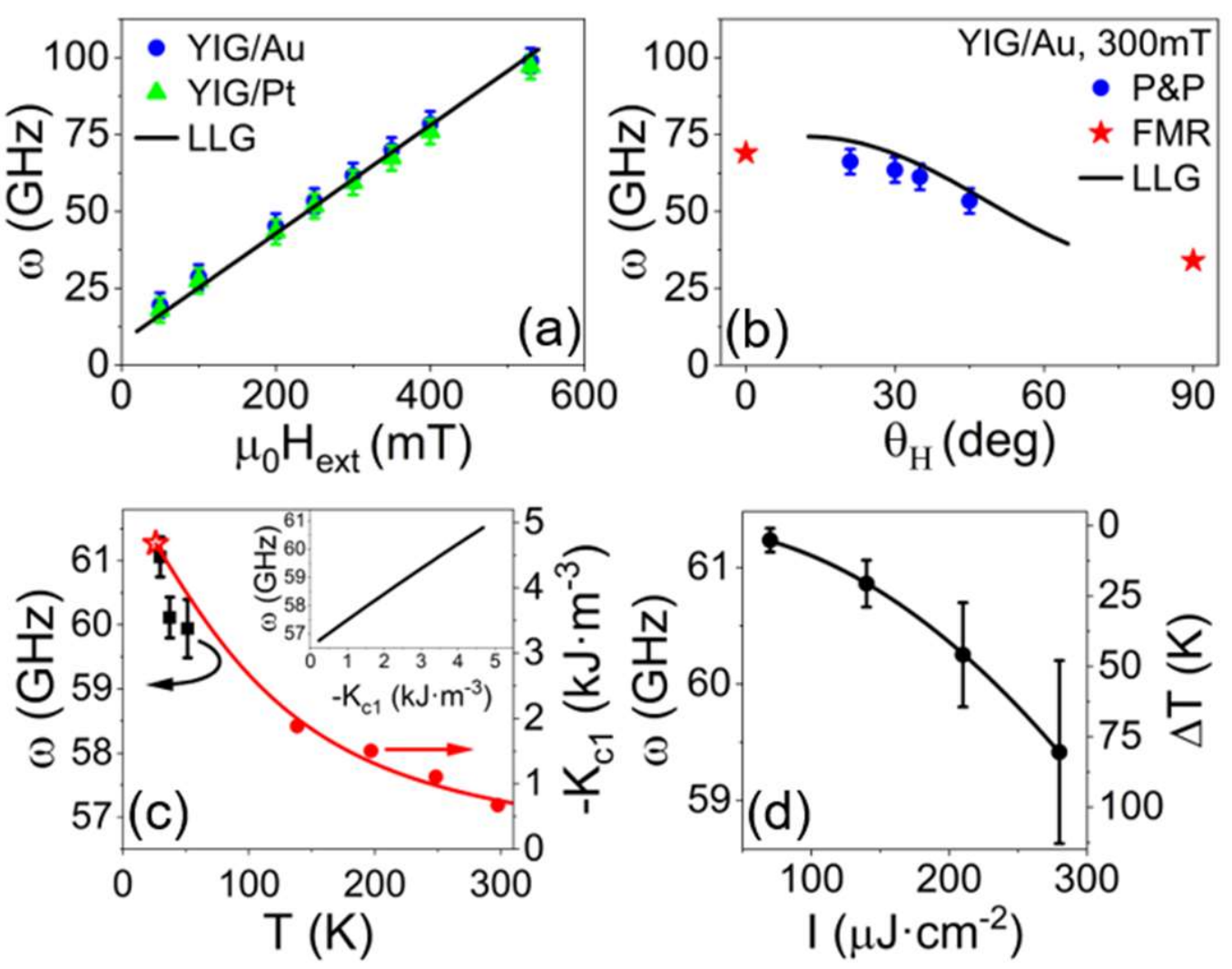

Fig. 2: (a) Frequency *f* of magnetization precession as a function of magnetic field applied at an angle $\theta_H$ = 40°, for YIG/Pt (blue dots) and YIG/Au (green triangles) at $T$ = 20 K and $I$ = 150 μJ/cm$^2$. The line is calculated from LLG equation (Eq. S3) with the free energy given by (Eq.S5) (b) Field-angle dependence of *f* in YIG/Au sample for $\mu_0 H_{ext}$ = 300 mT (blue dots), compared to a model by LLG model (line) and to frequencies measured by MW-FMR (red stars)[32]. (c) Temperature dependence of *f* in YIG/Au sample (black points), where $\mu_0 H_{ext}$ = 300 mT was applied at $\theta_H$ = 40°. The temperature dependence of cubic anisotropy constant $K_{c1}$ was obtained from Ref. 28 (red dots) and Ref. 32 (red star, T = 20 K). The data were fitted by an inverse polynomial dependence $K_{C1}(T) = \frac{1}{(a+bT+cT^2)}$, with parameters: a = 0.18 m$^2$/kJ; b= 9 x 10$^{-4}$ m$^2$/kJ.K; c = 9 x 10$^{-6}$m$^2$/kJ.K$^2$. Inset: Dependence $f(K_{c1})$ obtained from the LLG equation. (d) *f* as a function of pump pulse fluence *I*, from which the increase of sample temperature $\Delta T$ for the used pump fluences was evaluated using the *f(T)* dependence.

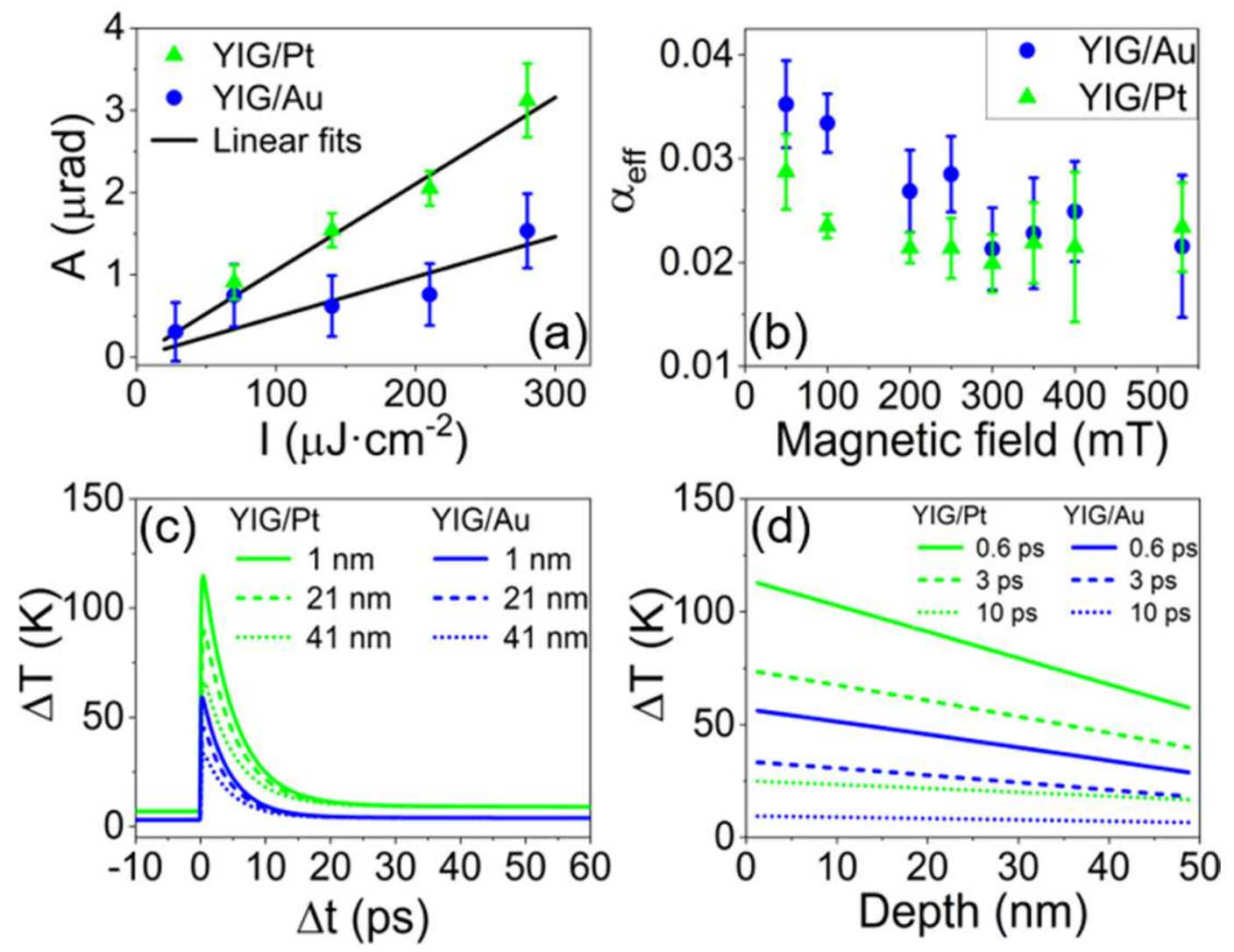

Fig. 3: Comparison of magnetization precession in YIG/Pt and YIG/Au samples. (a) Precession amplitude $A$ as a function of pump fluence $I$ (dots) with the corresponding linear fits $A = s \cdot I$. The parameter $s_{Pt}$ = (1.05 ±0.09)x$10^{-2}$ μrad.cm$^2$/μJ in the YIG/Pt, and $s_{Au}$ = (0.5±0.1)x$10^{-2}$ μrad.cm$^2$/μJ in YIG/Au. These dependencies were measured for $\mu_0 H_{ext}$ = 300 mT and $T_0$ = 20 K. In YIG/Pt sample the as-measured data obtained for $\theta_H$ = 40° are shown. In the YIG/Au sample, the $A$($I$) dependence was originally measured for $\theta_H$ = 21° and recalculated to $\theta_H$ = 40° according to the measured angular dependence, as described in detail in Supplementary Material, Section 3. (b) Gilbert damping $\alpha_{eff}$ for $H_{ext}$ applied at an angle $\theta_H$ = 40°. The values of $\alpha_{eff}$ result from fitting the TRMO signals to LLG equation; I = 140 μJ/cm$^2$. (c) and (d) Increase in lattice temperature as a function of time delay between pump and probe pulses for selected depths from the sample surface (c) and as a function of depth for fixed time delays (d). I = 140 μJ/cm$^2$, $T_0$ = 20 K. The heat capacities and conductivities of individual layers are provided in the Supplementary Material, Section 7.

# Thermally induced all-optical ferromagnetic resonance in thin YIG films: Supplementary Material

E. Schmoranzerová[1]*, J. Kimák[1], R. Schlitz[3] , S.T. B. Goennenwein[3], D. Kriegner[2,3], H. Reichlová[2,3,] , Z. Šobáň[2], G. Jakob[5], E.-J. Guo[5], M. Kläui[5], M. Münzenberg[4], P. Němec[1] , T. Ostatnický[1]

[1]Faculty of Mathematics and Physics, Charles University, Prague, 12116, Czech Republic

[2]Institute of Physics ASCR v.v.i , Prague, 162 53, Czech Republic

[3]Technical University Dresden, 01062 Dresden, Germany

[4]Institute of Physics, Ernst-Moritz-Arndt University, 17489, Greifswald, Germany

[5]Institute of Physics, Johannes Gutenberg University Mainz, 55099 Mainz, Germany

[6] Department of Physics, University of Konstanz, 78464 Konstanz, Germany

## 1. Magnetic characterization

### A. SQUID magnetometry

A superconducting quantum device magnetometer (SQUID) was used to characterize the magnetic properties of the thin YIG film at several sample temperatures. The magnetic hysteresis loops, detected with magnetic field applied in [2-1-1] crystallographic direction of the YIG layer, are shown in Fig. S1. As expected [t26], the saturation magnetization increases at low temperatures, which is accompanied by a slight increase in coercive field. At room temperature, the effective saturation magnetization is estimated to be $M_s$ = 95 kA/m. This value is in good agreement with the effective magnetization $M_{eff}$ obtained from the ferromagnetic resonance (FMR) measurement (see Section 1b), which indicates only a weak out-of-plane magnetic anisotropy [s1]. However, as discussed in detail in Ref. 26, the $M_s$ from our SQUID measurement is burdened by a relatively large error. Therefore, mere comparison of SQUID and FMR experiment is not sufficient to evaluate the size of the out-of-plane magnetic anisotropy. An additional experiment such as static magneto-optical measurement [28] is needed in order to get more precise estimation of the out-of plane magnetic anisotropy.

### B. FMR measurement

The SQUID magnetometry was complemented by so-called broad band ferromagnetic resonance measurements using a co-planar waveguide to apply electromagnetic radiation of a variable frequency $f = \omega/2\pi$ to the sample. The measurement was performed at room temperature and further details on the method can be found in Ref. s2. An exemplary set of spectra showing the normalized microwave transmission $|S_{21}|_{norm}$ obtained at different external fields magnitudes applied in the sample plane, is shown in Fig. S2(a). The set of Lorentzian-shape resonances can be fitted by the equation:

$$|S_{21}|_{\text{norm}} = \frac{B\left(\frac{\Delta\omega}{2}\right)^2}{\left(\frac{\omega}{2\pi}-\frac{\omega_0}{2\pi}\right)+\left(\frac{\Delta\omega}{2}\right)^2} + y_0 \quad \text{(S1)}$$

Where $f_0 = \omega_0/2\pi$ is the FMR resonance frequency, $\Delta\omega/2$ is the half width half maximum line width, $B$ the amplitude of the FMR line and $y_0$ a frequency independent offset. From an automated fitting of the set of lines obtained at different $H_{ext}$, we extract the magnetic field dependence of the resonance frequency $\omega_0/2\pi(H_{ext})$ [Fig. S2(b)] and linewidth $\Delta\omega(H_{ext})$ [Fig. S2(c)]. Clearly, the resonance frequencies correspond to the fundamental (Kittel) mode, and can correspondingly be fitted by the Kittel formula [s3]:

$$\frac{\omega_0}{2\pi} = \frac{\gamma}{2\pi}\sqrt{\mu_0 H_{\text{ext}}(\mu_0 H_{\text{ext}} + \mu_0 M_{\text{eff}})} \quad \text{(S2)}$$

Where $M_{eff}$ is the effective saturation magnetization that includes the out-of-plane anisotropy term, and $\gamma$ is gyromagnetic ratio. From this fit, it is possible to evaluate $M_{eff,\ Kittel}$ = 94.9 kA/m

From the linewidth dependence $\Delta\omega(H_{ext})=2\alpha\omega + \Delta\omega_0$ we can extract both the inhomogeneous line broadening and the Gilbert damping parameter, as shown in Fig. S2(c) [s2]. In our experiment, the inhomogeneous linewidth broadening is $\Delta\omega_0$ = 55.8 MHz, and the Gilbert damping parameter $\alpha$ = 0.001. Both values are on a higher side compared e.g. with YIG prepared by liquid phase epitaxy [s8] but in good agreement with typical YIG thin films similar to our layers, which were prepared by pulsed laser deposition [27]. This again confirms the good quality of the studied thin YIG films.

## 2. Processing of time-resolved magneto-optical data

In order to extract the parameters describing the precession of magnetization correctly from the time-resolved magneto-optical (TRMO) signals, it is first necessary to remove the slowly varying background on which the oscillatory signals are superimposed. For this purpose, we fitted the measured data by the second-order polynomial. The fitted curve was then subtracted from the measured signals, as demonstrated in Fig. (S3).

From the physical point of view, the background can be attributed to a slow return of magnetization to its equilibrium state after the pump beam induced heating, which can take place on the timescale of tens of nanoseconds [10]. Since both saturation magnetization $M_s$ and magnetocrystalline anisotropy $K_c$ are temperature-dependent, their temporal variation can in principle contribute to the background signal. However, as explained later in Section 4, the variation of $M_s$ is very weak at cryogenic temperatures. The heat-induced modification of $K_c$, and the resulting change of the magnetization quasi-equilibrium orientation, is, therefore, more probable origin of the slowly varying background, which is detected in the MO experiment by the Cotton-Mouton effect [28].

## 3. Angular dependence of precession amplitude

In order to mutually compare the values of precession amplitudes measured in YIG/Pt and YIG/Au samples at different angles of the external magnetic field $\theta_H$, it is necessary to correct their values for the value of $\theta_H$. The following procedure was used to correct the data presented in Fig. 3 of the main text.

First, we measured in detail angular dependence of the precession amplitude in the YIG/Au layer, which is presented in Fig. S4. Amplitude of the oscillatory signal detected in our experiment does not depend solely on the amplitude of the magnetization precession but also on the size of the magneto-optical (MO) effect. In our experimental setup, the change of $\theta_H$ was achieved by tilting the sample relative to the position of electromagnet poles [see Fig. 1(a)]. The MO response, however, varies also with the angle of incidence which is modified simultaneously with a change of $\theta_H$ [see Fig. 1(a)]. Therefore, it is not straightforward to describe the $A(\theta_H)$ analytically. Instead, we fitted the measured dependence $A(\theta_H)$ by a rational function in a form of $y = 1/(A+Bx^2)$, which is the lowest order polynomial function that can describe the signal properly. From the fit we derived a correction factor of 1.7 by which the amplitudes $A$ measured at $\theta_H = 21°$ has to be multiplied to correspond to that measured at $\theta_H = 40°$. This factor was then used to recalculate the intensity dependence of the precession amplitude $A(I)$ in YIG/Au measured at $\theta_H = 21°$ to the $A(I)$ at $\theta_H = 40°$, which could be directly compared to the $A(I)$ dependence detected at YIG/Pt for $\theta_H = 40°$ - see Fig. S4(b).

## 4. Temperature dependence of precession amplitude

In order to further investigate the origin of the laser-induced magnetization precession, the amplitude of the oscillatory MO signal was measured as a function of the sample temperature in YIG/Pt sample, see Fig. S5(a). In Fig. S5 (b), we show temperature dependence of saturation magnetization $M_s$, as obtained from Ref. 32

The only parameter changed within this experiment was the sample temperature. It is reasonable to expect that the size of the magneto-optical effect is not strongly temperature dependent in the studied temperature range between 20 and 50 K (see Ref. 28) Therefore, the dependence $A(T)$ presented in Fig S5 corresponds directly to the temperature dependence of magnetization precession amplitude. By comparing the $M_s(T)$ and $A(T)$ data, it is immediately apparent that the laser-induced heating would not modify $M_s$ enough to account for the large change of the magnetization precession amplitude with the sample temperature. Even assuming the most extreme laser-induced temperature increase $\Delta T \approx 80$ K shown in Fig. 3(c), the laser-induced $M_s$ variation would be less than 5%, while the precession amplitude changes by more than 50% between 20 and 50 K. In contrast, the magnetocrystalline anisotropy $K_{c1}$ changes drastically even in this relatively narrow temperature range [see Fig. 2(c)]. Consequently, the change of $K_{c1}$, which leads to a significant change of the position of quasi-equilibrium magnetization orientation in the studied sample (see Section 5) provides a more plausible explanation for the origin of the laser-induced magnetization precession in the YIG/metal layer.

## 5. LLG equation model

The data were modelled by numerical solution of the Landau-Lifshitz-Gilbert (LLG) equation, as defined in [s9]:

$$\frac{d\boldsymbol{M}(t)}{dt} = -\mu_0 \gamma \left[\boldsymbol{M}(t) \times \boldsymbol{H_{eff}}(t)\right] + \frac{\alpha}{M_s}\left[\boldsymbol{M}(t) \times \frac{d\boldsymbol{M}(t)}{dt}\right], \text{(S3)}$$

where $\gamma$ is the gyromagnetic ratio, $\alpha$ is the Gilbert damping constant, and $M_S$ is saturated magnetization.The effective magnetic field $H_{eff}$ is given by:

$$\boldsymbol{H_{eff}}(t) = \frac{\partial F}{\partial \boldsymbol{M}} \qquad \text{(S4)}$$

where $F$ is energy density functional that contains contributions from the external magnetic field $H_{ext}$, demagnetizing field and the magnetic anisotropy of the sample. We consider the form of $F$ including first- and second-order cubic terms as defined in Ref. [t24]. The polar angle $\theta$ is measured with respect to the crystallographic axis [111] and the azimuthal angle $\varphi$ = 0 corresponds to the direction $[2\bar{1}\bar{1}]$, with an appropriate index referring to the magnetization position (index $M$) or the direction of the external magnetic field (index $H$). The resulting functional takes the form (in the SI units):

$$F = -\mu_0 HM[\sin\theta_M \sin\theta_H + \cos\theta_M \cos\theta_H \cos(\varphi_H - \varphi_M)] + \left(\tfrac{1}{2}\mu_0 M^2 - K_\mathrm{u}\right)\sin^2\theta_M + \frac{K_\mathrm{c1}}{12}\left[7\cos^4\theta_M - 8\cos^2\theta_M + 4 - 4\sqrt{2}\cos^3\theta_M \sin\theta_M \cos 3\varphi_M\right] + \frac{K_\mathrm{c2}}{108}\left[-24\cos^6\theta_M + 45\cos^4\theta_M - 24\cos^2\theta_M + 4 - 2\sqrt{2}\cos^3\theta_M \sin\theta_M (5\cos^2\theta_M - 2)\cos 3\varphi_M + \cos^6\theta_M \cos 6\varphi_M\right], \qquad \text{(S5)}$$

where $\mu_0$ is the vacuum permeability and we consider the following values of constants: magnetization $M$ = 174 kA/m, first-order cubic anisotropy constant $K_{c1}$=–4.68 kJ/m$^3$, second-order cubic anisotropy constant $K_{c2}$ = –222 J/m$^3$ [t24].

For modelling the dependence of precession frequency on the external magnetic field $H_{ext}$ [Fig. 2 (a)] and on the angle $\theta_H$ [Fig. 2 (a)], we assumed that in a steady state magnetization direction is parallel to $H_{ext}$, i.e. $\theta_M = \theta_H$, and $\varphi_M = \varphi_H$. This is surely fulfilled for large enough magnitude of $\mathbf{H_{ext}}$. Since the coercive field is very small, we can assume the procedure to be correct. Further correspondence to experimental data

Evaluation of the Gilbert damping factor from the as-measured magneto-optical oscillatory data was done by fitting signals by a theoretical curve calculated by solving numerically LLG equation [Eq. (S3)]. We considered the magnetization free energy density in a form of Eq. (S5) using magnitude and direction of the external magnetic field from the experiment. The electron g-factor was set to 2.0 and then the Gilbert factor and five parameters of the fourth-order polynomial to remove the background MO signal were the fitting parameters. The resulting dependence of fitted effective Gilbert factors $\alpha_{eff}$ on external magnetic field is displayed in Fig. 3(b) in the main text, from which the field-independent Gilbert factor $\alpha$ can be evaluated.

## 6. Comparison of Gilbert damping parameter from MW-FMR and TRMO experiments

The Gilbert damping from the room-temperature FMR measurement on the YIG film $\alpha \approx 1\cdot10^{-3}$ and the results from fits of the low-temperature pump&probe data $\alpha \approx 2\cdot10^{-2}$, differ by an order of magnitude. As detailed in the main text, we attribute this difference to the different sample temperatures in the AO-FMR and MW-FMR measurements. However, one might also argue that the increased damping in the optical experiments is caused either by a spatial inhomogeneity of the magnetization oscillations or it is the result of the perturbation of the YIG surface.

In the former case, we expect that the spatial inhomogeneity of the temperature distribution [see Fig. 3(d)] causes the magnetization to oscillate in a form of a superposition of harmonic waves with well-defined in-plane wavevectors. Considering the dispersion of the allowed oscillatory modes [s4] and including the relevant value of the exchange stiffness [s5], we revealed that neither the inhomogeneity due to the finite cross section of the excitation laser beam nor the temperature gradient perpendicular to the sample surface can cause such a strong decrease of the Gilbert damping factor that is observed experimentally. Here, we provide an estimate on which time scales the mode dispersion influences the decay of the signal if the exchange stiffness is taken into account. Following [s5], the mode dispersion is described by the additive exchange field in the form:

$$\mu_0 H_{\mathrm{ex}} = D\left[\frac{\pi^2}{d^2}n^2 + k_{\parallel}^2\right],$$

where $D \approx 5\cdot10^{-17}$ T.m$^2$ is the exchange stiffness, $n$ is the order of the confined magnon mode, $d$ is the YIG layer thickness and $k_{\parallel}$ is the in-plane magnon wave vector. We consider here only the $n$ = 0 case since this is the only visible harmonic mode observed in the experimental MO data, as proven by the numerical fitting. Note that the frequency shift $\Delta\omega/2\pi = |\gamma|\mu_0 H_{\mathrm{ex}}/2\pi$, where the symbol $\gamma$ stands for the electron gyromagnetic ratio, of the $n$ = 1 mode would be 5.5 GHz, which would be then clearly distinguishable from the basic $n$ = 0 mode in the lowest external magnetic fields. The in-plane wave vector $k_{\parallel}$ can be calculated from the FWHM (full width at half maximum) width of the laser spot on the sample $L$, which is about 30 µm in our case, that leads to the order of magnitude $k_{\parallel} \approx (2\pi/L) \approx 10^5$ m$^{-1}$. The frequency increase due to the finite laser spot size can be estimated as $\Delta\omega/2\pi = |\gamma|Dk_{\parallel}^2/2\pi \approx$14 kHz. Inverse of this value ($\approx$ 0.1 ms) determines the typical time scale at which the magnon dynamics is influenced by their dispersion due to the finite laser spot size, which is clearly out of the range of the experimental time scale.

The presence of a metallic layer on the top of the YIG sample surface can result into two significant damping processes. First, the magnetization oscillations (and thus oscillations of the macroscopic magnetic field) are coupled to electromagnetic modes which penetrate the surrounding material and can be eventually radiative for small magnon wave vectors. Penetration into conductive material in turn causes energy dissipation through finite conductivity of such material. We checked the magnitudes of the additional damping caused by the radiative field and energy dissipation in a thin metallic layer and we found that these processes exist but the additional energy loss cannot explain the observed magnitude of the Gilbert damping parameter. The second possible explanation of the increased precession damping due to the presence of the metal/YIG surface may be that there is an additional perturbation to otherwise homogeneous sample due to some inhomogeneity through surface roughness or spatially inhomogeneous local spin pinning. Since both the surface roughness and spin pinning can depend on the composition of the capping layer, it can also cause a minor difference in the resulting damping factor, as observed in Fig. 3(b).

Overall, we attribute the experimentally observed difference in Gilbert damping measured by FMR and pump&probe techniques to the difference in ambient temperatures that were used in these experiments, which is in accord with the results of Ref. t22.

## 7. Heat propagation in YIG/Pt and YIG/Au

Heat propagation in our sample structures was modelled in terms of the heat equation:

$$\frac{\partial T}{\partial t} = \frac{\lambda}{c} \Delta T \, , \qquad \text{(S4)}$$

where $T$ is the local temperature, $\lambda$ is the local thermal conductivity, $c$ is the heat capacity and the symbol $\Delta$ denotes the Laplace operator. The spatio-temporal temperature distribution in the studied sample has been calculated by a direct integration of Eq. (S4) in a time domain, assuming excitation of the metallic layer by an ultrashort optical pulse [with a temporal duration of 100 fs (FWHM)]. We have taken the whole structure profile of vacuum/metal/YIG/GGG into consideration, assuming that the GGG substrate had a perfect heat contact with the cold finger of the cryostat, which has been held on a constant temperature. The respective heat conductivities ($\lambda$) and heat capacities ($c$) were set to the following values. Au: $\lambda$ = 5 W/m K [s6], $c = 1.3 \cdot 10^4$ J/cm$^3$, Pt: $\lambda$ = 10 W/m K [s7], $c = 1.2 \cdot 10^4$ J/cm$^3$, YIG: $\lambda$ = 60 W/m K, $c = 6.7 \cdot 10^3$ J/cm$^3$, GGG: $\lambda$ = 300 W/m K, $c = 2.1 \cdot 10^4$ J/m$^3$.

To evaluate the initial heat transfer from the optical pulses to the capping metallic layer, we considered the proper geometry of our experiment, i.e. a 8 nm thick metallic layer deposited on the YIG sample, the incidence angle of the laser beam of 45 degrees and its *p*-polarization. We then used optical constants of gold and platinum in order to calculate transmission and reflection coefficients of a nanometer-thick metallic layers by means of the transfer matrix method. From those, we estimated the efficiency of power conversion from the optical field to heat to be 3% for gold and 6.5% for platinum. The total amount of heat density was then calculated by multiplication of the pump pulse energy density and the above-mentioned efficiency.

The data shown in Fig. 3(c)-(d) were then extracted from the full spatio-temporal temperature distribution. Clearly, the temperature increase in the YIG/Pt sample is approximately twice larger than that of the YIG/Au sample as a consequence of twice larger efficiency of the light-heat energy conversion in favour of platinum. Correspondingly, also the amplitudes of the MO oscillations in Fig. 3(a) reveal the ratio $\approx$ 2:1.

**FIGURES**

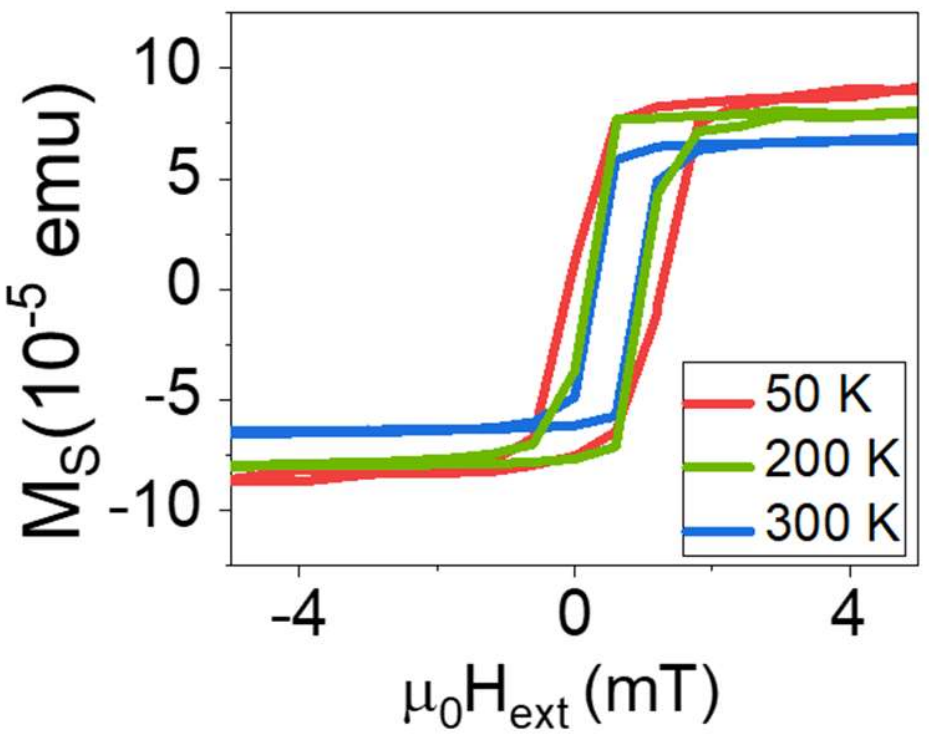

Fig. S1: Magnetic hysteresis loops measured by SQUID magnetometry with magnetic field $H_{ext}$ applied in direction [2-1-1] at several sample temperatures. The saturation magnetization obtained from SQUID magnetometry measurement at room temperature is roughly $M_s = 95\ kA/m$, assuming a YIG layer thickness of 50 nm.

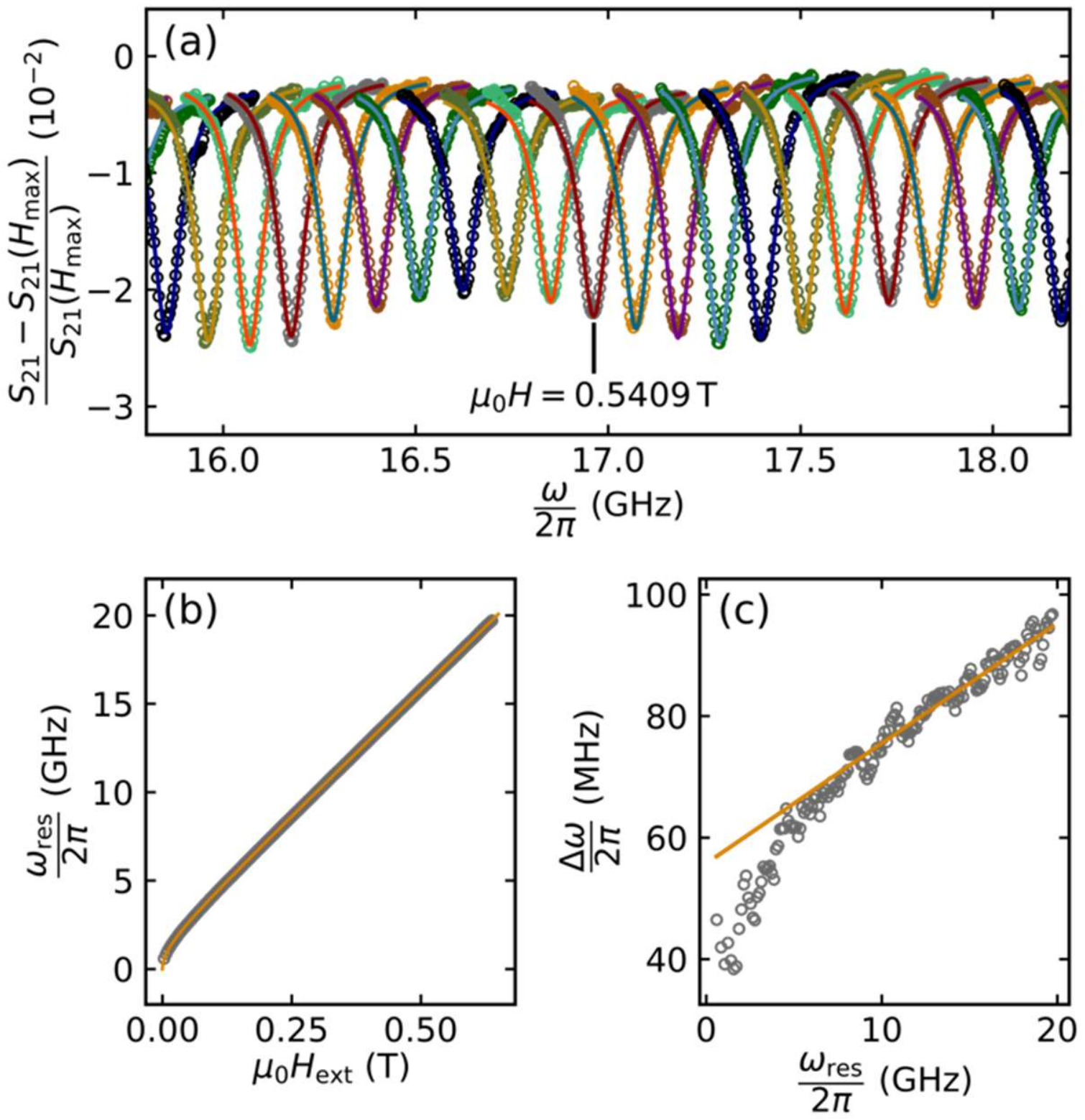

Fig. S2: (a) Ferromagnetic resonance spectra measured at room temperature for several different external magnetic field magnitudes $\mu_0 H_{ext}$ from 0 to 540 mT applied in the sample plane. Resonance peaks were fitted by Eq. (S1) and the obtained resonance frequencies and linewidths are plotted as points in panels (b) and (c), respectively. The lines correspond to fit by Kittel formula [Eq. (S2)], which enables to evaluate effective magnetization $M_{eff}$ = 94.9 kA/m and Gilbert damping parameter of $\alpha$ = 0.001.

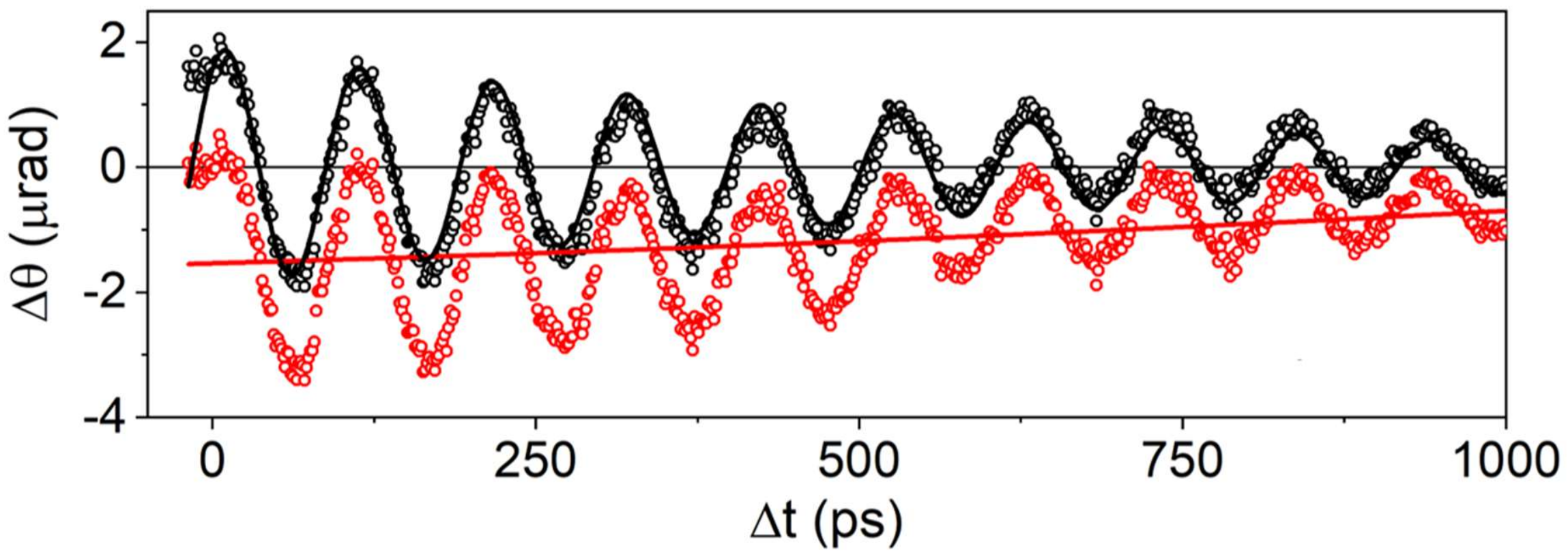

Fig. S3: Removal of slowly varying background from time-resolved magneto-optical signals. The red dots correspond to as-measured signals, line indicates the polynomial background that is subtracted from the raw signals. Black dots show the signal after background subtraction, black line representing the fit by Eq. (1) of the main text. The data were taken at external field of $\mu_0 H_{ext}$ = 300 mT, temperature 20 K and pump fluence $I$ = 140 μJ/cm$^{2}$..

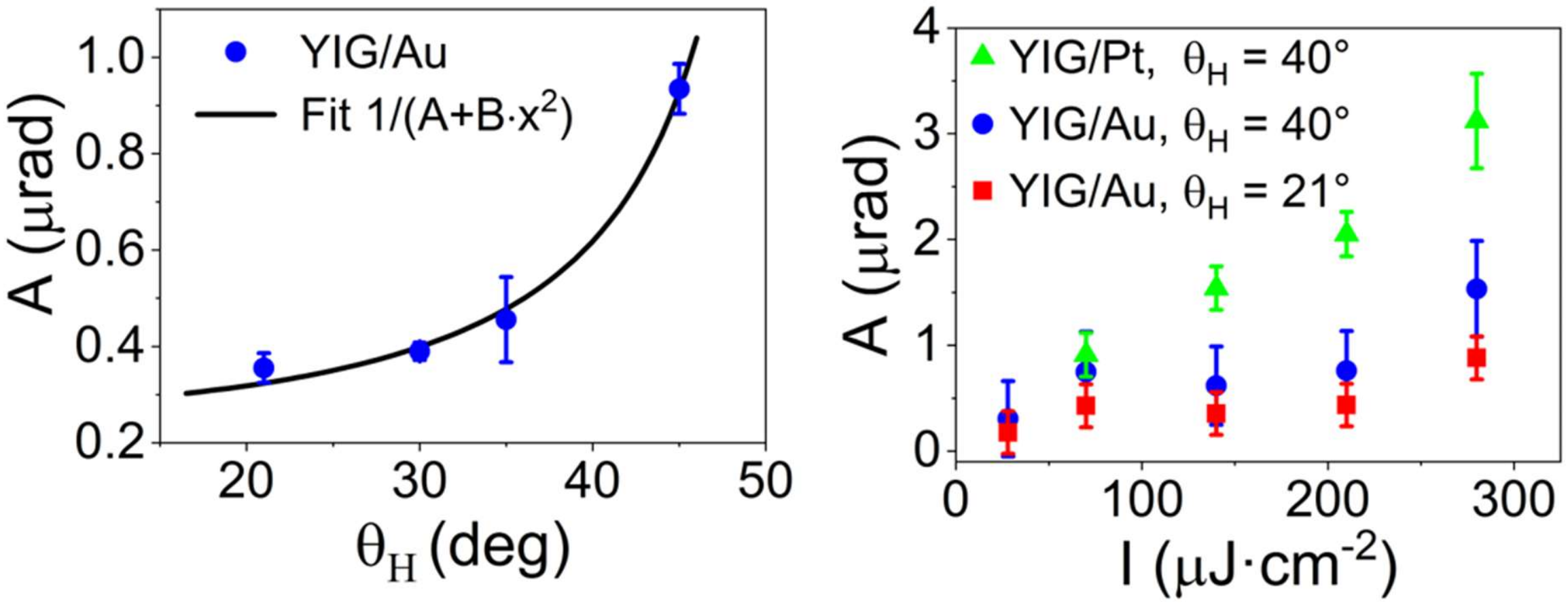

Fig. S4: (a) Dependence of the amplitude *A* of oscillatory magneto-optical signal on the sample tilt (different field angles of magnetic field $\theta_H$ ) measured in YIG/Au sample. $\mu_0 H_{ext}$ = 300 mT, temperature *T* = 20 K and pump fluence *I* = 150 μJ/cm$^{2}$.. (b) Pump intensity dependence of *A* measured for YIG/Au sample at $\theta_H$ =21° (red points), the same dependence recalculated to correspond to $\theta_H$ =40 (blue points) where *A(I)* was measured for YIG/Pt sample (green points); *T* = 20 K.

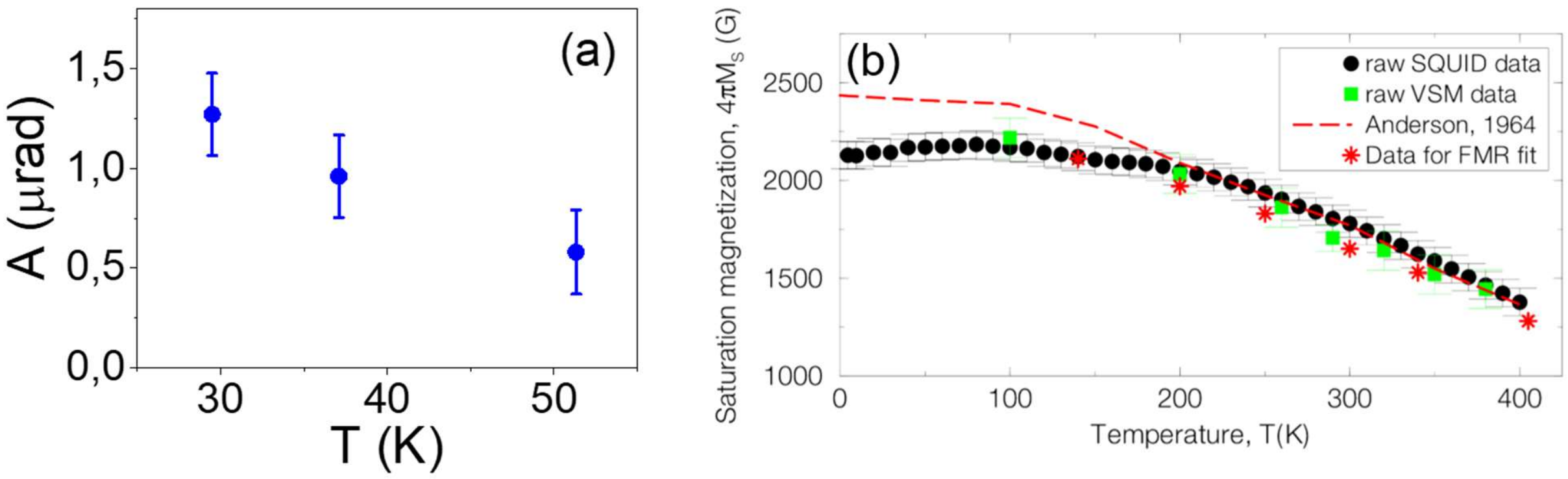

Fig. S5: (a) Temperature dependence of amplitude of the time-resolved magneto-optical signals measured for external field $\mu_0 H_{ext}$ = 300 mT applied at an angle $\theta_H$ = 30°. (b) Temperature dependence of saturation magnetization $M_s$ obtained from Ref. 32.